\documentclass[a4paper,11pt]{article}
\usepackage{pos}
\renewcommand{\logo}{\relax}

\title{Generalization of lattice Dirac operator index}

\author[a]{Shoto Aoki}
\author[b]{Hajime Fujita}
\author*[c]{Hidenori Fukaya}
\author[d]{Mikio Furuta}
\author[e]{Shinichiroh Matsuo}
\author[c]{Tetsuya Onogi}
\author[c]{Satoshi Yamaguchi}

\affiliation[a]{
Interdisciplinary Theoretical and Mathematical Sciences Program (iTHEMS),
  RIKEN,
  Wako, Japan
}
\affiliation[b]{
Faculty of Science, Japan Women’s University, Mejirodai, Bunkyo-ku, Tokyo 112-8681, Japan
}
\affiliation[c]{Department of Physics, Osaka University, 
        Toyonaka, Osaka 560-0043 Japan}

\affiliation[d]{Graduate School of Mathematical Sciences, The University of Tokyo, Komaba, Meguro-ku, Tokyo 153-8902, Japan}
\affiliation[e]{Graduate School of Mathematics, Nagoya University, Nagoya, Japan}

\abstract{
  We provide a comprehensive lattice formulation of various types of the Dirac operator indices,
  employing $K$-theory to classify the Wilson Dirac operator via its spectral flow. In contrast to the index of the overlap Dirac operator defined through the Ginsparg-Wilson relation, which is restricted to flat tori in even dimensions, our formulation offers several key advantages: 1) It can be applied straightforwardly to the Atiyah-Patodi-Singer index for manifolds with boundary. 2) The boundary can be curved, allowing for the inclusion of gravitational background effects. 3) The mod-2 index in both even and odd dimensions can be defined as a natural extension of the same formulation. In this talk, we present the mathematical proof and provide numerical evidence supporting the formulation.

OU-HET-1304
}

\FullConference{The 42nd International Symposium on Lattice Field Theory (LATTICE2025)\\
2-8 November 2025\\
Tata Institute of Fundamental Research, Mumbai, India\\}

\begin{document}
\maketitle

\section{Introduction}

To understand topology of gauge fields on
the lattice, the Atiyah-Singer(AS) index \cite{Atiyah:1968mp} defined by 
the overlap Dirac operator \cite{Neuberger:1997fp}
(as well as of the perfect action \cite{Hasenfratz:1998ri})
has played a key role.
The overlap Dirac operator is defined by
\begin{equation}
  \label{eq:ovdef}
D_{\rm ov} = \frac{1}{a}\left(1+\gamma_5{\rm sgn}(H_W)\right),
\end{equation}
where $H_W=\gamma_5(D_W-M)$ is the Wilson Dirac operator 
with a negative mass we  take $M=1/a$ as a standard choice.
Since this operator satisfies the Ginsparg-Wilson(GW) relation \cite{Ginsparg:1981bj},
\begin{equation}
\Gamma_5 D_{\rm ov}+D_{\rm ov}\Gamma_5= 0,
\end{equation}
where $\Gamma=\gamma_5(1-aD_{\rm ov}/2)$ is the modified
chirality operator, it realizes an exact chiral symmetry on the lattice \cite{Luscher:1998pqa}.
Moreover, the AS index 
\begin{equation}
 {\rm Tr}\;\Gamma_5 = n_+-n_-,
\end{equation}
is well-defined, where $n_\pm$ denotes the number of zero modes with $\pm$ chirality,
which correctly reproduces the topological charge in the continuum limit.

We note, however, that the index defined above can be
written by the massive Wilson Dirac operator only:
\begin{equation}
\label{eq:TrGamma5}
  {\rm Tr}\;\Gamma_5 = -\frac{1}{2}{\rm Tr}\;{\rm sgn}(H_W) = -\frac{1}{2}\sum_{\lambda_W}{\rm sgn}\lambda_W=:
  -\frac{1}{2} \eta(H_W),
\end{equation}
where $\eta(H_W)$ is known as the $\eta$ invariant in mathematics, 
which was introduced by Atiyah, Patodi and Singer \cite{MR397797}.
Here we denote the real eigenvalues of $H_W$ by $\lambda_W$.

The above equality is not a coincidence.
In \cite{Aoki:2024sjc}, we gave a mathematical proof that
a one parameter family of $H_W$ can be identified as an element in $K$-theory
and reproduces the continuum Dirac operator index for sufficiently small lattice spacings.
Note that our formulation does not require the GW relation at all.
Namely, the Wilson Dirac operator is good enough to describe the gauge field topology.
For different mathematical approaches using Wilson Dirac operators,
we refer the readers to \cite{MR4275791,MR4407739}.
Another extension to the mod 8 topological invariant is tried in \cite{Araki:2025xly, Araki}.

Moreover, $K$-theory can generalize this formula to various types of index 
where the GW relation is difficult or absent\footnote{
We refer the readers to \cite{Clancy:2023ino}, in which
they try generalizing of the GW relation.
}.
In this talk, we will show that the Wilson Dirac operator can
reproduce the index in the following cases: 1)
Atiyah-Patodi-Singer(APS) index \cite{MR397797} on a manifold with boundary,
2) the boundary is curved \cite{Aoki:2022aez},
which induces nontrivial gravitational background,
and 3) mod-two index in odd or even dimensions.

\section{Massless vs. massive Dirac operators in $K$-theory}

First, let us consider continuum theory and
compare the massless and massive Dirac operators in terms of $K$-theory.
$K$-theory is a generalized cohomology theory,
which is useful in classifying the vector bundles.
When we compare the two vector bundles $V_1$ or $V_2$
over a same base space $X$, a linear operator $D_{21}: V_1\to V_2$
and its conjugate $D_{12}=-D_{21}^\dagger : V_2\to V_1$ are essential.
Namely,
\begin{equation}
  D = \left(
  \begin{array}{cc}
    0 & D_{12}\\
    D_{21} & 0
  \end{array}  
  \right),
\end{equation}  
which anticommutes with $\gamma=\mathrm{diag}(1,-1)$,
gives an element of the group called $K^0(X)$.

If we set $V_1$ as a left-handed fermion field and
$V_2$ as a right-handed fermion field,
the massless Dirac operator $D$
determines an element of $K^0(X)$.
When we are interested in global structure only,
we can forget about details of the base manifold $X$ by taking 
the so-called $K$-theory pushforward,
$G: K^0(X) \to K^0(\mathrm{point})\cong \mathbb{Z}$,
where ``$\mathrm{point}$'' is an abstract zero-dimensional point.
Note that this map corresponds to 
``integration'' over $X$ in the standard cohomology theory.
With this map, a lot of information is lost but
one topological invariant remains, which is the Dirac operator index.

It is known that the group $K^0(\mathrm{point})$ is
isomorphic to another group $K^1(I,\partial I)$, where
$I$ denotes an interval where its two end points are
denoted by $\partial I$.
This is known as the suspension isomorphism.
Here and in the following, we parametrize it as $I=[-1,1]$.
The elements of $K^1(I,\partial I)$  are given by
the one-parameter family of Dirac operators
$H_s$ for $s\in I$, under a condition that
$H_s$ has no zero mode at $s=\pm 1$.
Unlike the elements of $K^0(\mathrm{point})$,
$H_s$ is not required to anticommute with the chirality operator $\gamma$,
which is reflected to change of the superscript from ``0'' to ``1''.
The elements of $K^1(I,\partial I)$ are classified by
the spectral flow of $H_s$ along $s\in I$, which
coincides with the Dirac operator index.
By identifying $H_s$ as the massive Dirac operator,
we can express the original Dirac operator index in
terms of the massive operator.

Let us explicitly demonstrate it in continuum theory considering 
the massive Dirac operator multiplied by $\gamma$:
\begin{equation}
H_s=\gamma (D+ms),\;\;\; s\in [-1,1],
\end{equation}
where we set $m>0$.
For the zero modes of the original massless Dirac operator $D$,
they remain the eigenstates of $H_s$ with the eigenvalues $\pm ms$,
where the $\pm$ sign equals to the chirality.
For nonzero modes of $D$, we have an anticommutation relation
$\{D, H_s\}=0$, which indicates that
for any eigenstate $H_s\phi = \lambda_s \phi$,
there exists another eigenstate $D\phi$ satisfying $H_s D\phi=-\lambda_s \phi$,
to form a pair $\pm \lambda_s$, unless $D\phi=0$.
The paired eigenvalues are explicitly given by  $\pm \sqrt{\lambda_0^2+m^2s^2}$,
which never cross zero.

\begin{figure*}[tbhp]
  \centering
  \includegraphics[width=0.45\columnwidth]{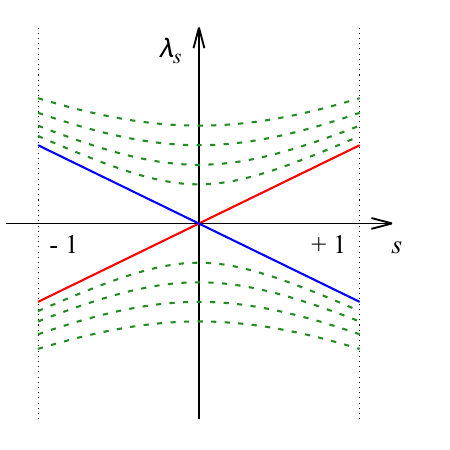}
\includegraphics[width=0.45\columnwidth]{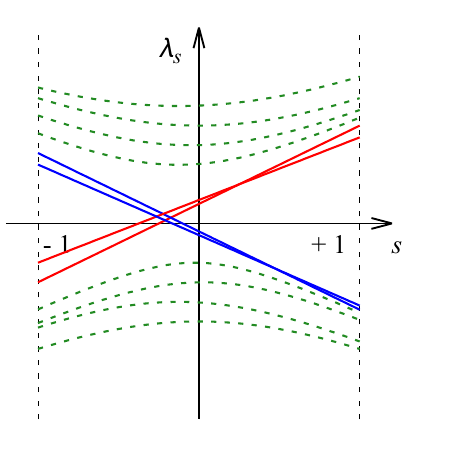}
  \caption{
    Left panel : continuum eigenvalue spectrum of the massive Dirac operator where the mass term is 
varied by $-sm$ with $s\in [-1,1]$. 
    Right panel : an example of deformed spectrum by chiral symmetry breaking.
  }
  \label{fig:sfcont}
\end{figure*}

In the left panel of Fig.~\ref{fig:sfcont}, we illustrate the spectrum of the massive Dirac operator
as a function of $s\in [-1,1]$.
The nonzero modes make parabola-like curves, which is $\pm$ symmetric.
On the other hand, the chiral modes cross zero from negative to positive when the chirality is $+$
and go from positive to negative when they have the $-$ chirality.
If we count the net zero-crossing lines, which is the so-called spectral flow 
(we denote by $\mathrm{sf}[H_s]$) which gives an element of $K^1(I,\partial I)$,
it is obvious that the spectral flow agrees the original Dirac operator index 
$\mathrm{Ind}[D] =n_+-n_-$.
Here we may say that we count the standard Dirac index in terms of $K^0(\mathrm{point})$
by the ``points'' at $m=0$, while the spectral flow for $K^1(I,\partial I)$
is counted by ``lines'' which cross zero.

The spectral flow can also be expressed by the $\eta$ invariant
\begin{equation}
 \mathrm{Ind}[D] = -\frac{1}{2}\eta(H_1) +\frac{1}{2}\eta(H_{-1}).
\end{equation}  
Note that whenever a eigenvalues crosses zero, $\eta(H_s)$ jumps by two.
Therefore, the $\eta$ invariant of the massive Dirac operator equals
to the index of the massless Dirac operator,
which is guaranteed by the suspension isomorphism.
It is interesting that a similar formula is found on the lattice in
(\ref{eq:TrGamma5}) but with the overlap Dirac index on the
left hand side, while the right hand side is the $\eta$ invariant
of the Wilson Dirac operator.

The key observation here is that 
the group $K^1(I,\partial I)$ does not require 
the chiral symmetry, and it is easier to consider on a lattice.
When the chiral symmetry is broken, 
the Dirac spectrum would be deformed like the right panel of Fig.~\ref{fig:sfcont}.  
The standard definition of the Dirac index is difficult
now since both of the massless point or zero mode 
are no longer well-defined.
However, the spectral flow, counting the lines crossing zero,
is still well-defined as far as there are gap from zero at $s=\pm 1$.
Counting the ``lines'' is easier and more stable than counting the ``points''\footnote{
It was proposed in \cite{Itoh:1987iy} that the index can be evaluated by 
the spectral flow of the Wilson Dirac operator 
even before the overlap Dirac operator was known.
But its mathematical meaning related to $K$-theory was not discussed.
See also \cite{Adams:1998eg}.
}.

\section{Main theorem and its generalization}

In Ref.~\cite{Aoki:2024sjc}, we proved
that the spectral flow of the massive Wilson Dirac operator 
on an even-dimensional flat periodic square lattice,
for a sufficiently large $m>0$
equals to the continuum Dirac index at sufficiently small lattice spacings:
\begin{equation}
\mathrm{Ind}[D] = \mathrm{sf}[\gamma(D_W-s m)] =-\frac{1}{2}\eta(\gamma(D_W-m)).
\end{equation}
Note that $\frac{1}{2}\eta(\gamma(D_W+m))$ is trivially zero and
it is thus neglected in the right hand side.
In the proof, we showed that $\mathrm{sf}[\gamma(D_W-s m)]$
gives a well-defined element of $K^1(I,\partial I)$
and coincides with $\mathrm{sf}[\gamma(D-s m)]$ in the continuum theory.  
We would like to stress that in the proof we did not rely on the GW relation at all.
The $\eta$ invariant itself is a well-defined mathematical object,
and its equivalence to the index of the overlap Dirac operator is just a by-product.

Recently in \cite{Aoki:2025gca,Aoki:2026mgx}, we extended the formulation 
to the cases where the mass term $m\epsilon$ at $s=1$ has domain-walls
\cite{Kaplan:1992bt, Shamir:1993zy}
where 
\begin{equation}
 \epsilon(x)=\left\{
\begin{array}{cc}
 +1 & \mathrm{for}\;x \in X_+\\
 -1 & \mathrm{for}\;x \in X_-
\end{array}
\right.,
\end{equation}
where $X_\pm$ are subregions of the even-dimensional periodic square lattice.
In this case, the corresponding spectral flow of the continuum domain-wall Dirac operator
$\mathrm{sf}\left[\gamma\left(D-\frac{s+1}{2} m \epsilon -\frac{s-1}{2}m\right)\right]$ was
prove to be equal to the APS index on $X_+$ \cite{Fukaya:2017tsq,MR4179728, Fukaya:2019myi}
(see also a summary review \cite{Fukaya:2021sea} and other lattice applications \cite{Nguyen:2024wck}).
Namely, we have proved for sufficiently small lattice spacings that
\begin{equation}
\mathrm{Ind}_\mathrm{APS}[D|_{X_+}] = \mathrm{sf}\left[\gamma\left(D_W-\frac{s+1}{2} m \epsilon -\frac{s-1}{2}m\right)\right] =-\frac{1}{2}\eta(\gamma(D_W-m \epsilon)),
\end{equation}
for sufficiently large $m>0$.
The APS index $\mathrm{Ind}_\mathrm{APS}[D|_{X_+}]$ in the continuum theory
is defined on the submanifold $X_+$ with the boundary located at the domain-wall
between $X_+$ and $X_-$ where a non-local boundary condition called APS condition
is imposed.
In our formulation with the domain-wall fermions, 
we use the whole lattice, which is just a torus in the continuum limit,
to avoid the non-local boundary condition.
In contrast to the standard AS index, 
the APS index has no counterpart by the overlap Dirac operator,
since the GW relation is difficult to maintain 
with any boundary conditions \cite{Luscher:2006df}.
The proof straightforwardly goes to show that the 
spectral flow of the lattice domain-wall fermion Dirac operator
 $\mathrm{sf}\left[\gamma\left(D_W-\frac{s+1}{2} m \epsilon -\frac{s-1}{2}m\right)\right]$
is a well defined element of $K^1(I,\partial I)$ with an additional assumption that
the domain-wall operator has no zero mode at $t=1$.

Our formulations via $K$-theory is so robust that
application to more general fermion systems with various symmetries
in general dimensions, is straightforward.
An important application is the mod-two AS and APS indices
where the Dirac operator is real.
In this case, the continuum Dirac operator $D$ has
a $\pm$ symmetric spectrum, since every nonzero eigenmode $\phi$
with an eigenvalue $\lambda$,
satisfying $D\phi = i\lambda \phi$, makes a pair with $\phi^*$,
which satisfies $D\phi^* = -i \lambda \phi^*$.
The number of zero modes of $D$, is a topological invariant, called
the mod-two index.

In $K$-theory, the mod-two index on a closed manifold
gives an element in the group $KO^{-1}(\mathrm{point})$,
which classifies the real vector bundles.
Here we consider general dimensions so that we do not assume existence of the chirality operator,
which is reflected in the superscript ``$-1$''\footnote{
The minus sign comes from the fact that $D$ is anti-Hermitian, rather than Hermitian.
}.

By the suspension isomorphism: $KO^{-1}(\mathrm{point})\cong KO^0(I,\partial I)$,
we can express the mod-two index by the mod-two spectral flow of the
massive Dirac operators \cite{Carey2016SpectralFF}.
It was also shown that the mod-two spectral flow of the domain-wall
fermion Dirac operators equals to the mod-two APS index on $X_+$. 
When there is a chirality(-like) operator $\gamma$,
which anticommutes with $D$, $H_s=\gamma\left(D-\frac{s+1}{2} m \epsilon -\frac{s-1}{2}m\right)$ gives an element of $KO^0(I,\partial I)$.
In general odd dimensions, however, the simple massive Dirac operator $D+m$
is neither Hermitian or anti-Hermitian.
Therefore, when the $\gamma$ is absent,
we ``double'' the fermion flavors to introduce
the family of real Hermitian operators:
\begin{equation}
  H_s = i\tau_2\otimes D - \tau_1\otimes \left(\frac{s+1}{2} m \epsilon
  +\frac{s-1}{2}m\right),
\end{equation}
where $\tau_{1,2}$ are Pauli matrices, which anticommutes with $\tau_3$.
In either case, every eigenvalues of $H_s$ make $\pm$ pairs
and the standard spectral flow is  always zero.
We thus count the mod-two spectral flow $\mathrm{sf}^{\mathrm{mod-2}}\left[H_s\right]$
by the number of zero-crossing pairs modulo two \cite{Fukaya:2020tjk}.

In \cite{Aoki:2026mgx},
including the domain-wall mass term, we have proved that
for a sufficiently large mass $m>0$,
\begin{equation}
  \mathrm{Ind}^{\mathrm{mod-2}}_\mathrm{APS}[D|_{X_+}]
  = \mathrm{sf}^{\mathrm{mod-2}}\left[(H_W)_s\right],
\end{equation}
holds at sufficiently small lattice spacings,
where $(H_W)_s$ is the lattice version of $H_s$ just by replacing $D$
with the Wilson Dirac operator $D_W$.

As described above, the spectral flow of the massive Wilson Dirac operator
gives a unified description of various types of the Dirac operator index.
In our formulation, the chiral symmetry, or the GW relation, is not necessary at all
(besides, it agrees with the overlap index on periodic lattices).
Boundaries can be introduced by domain-walls, avoiding imposing the nonlocal
APS boundary conditions.
The domain-walls can be flat or curved to include gravitational background \cite{Aoki:2022aez}.
The standard and mod-two versions in arbitrary dimensions are  treated in a unified way.

\section{Numerical evidences}

Now we would like to show some numerical evidence that
the spectral flow of the Wilson Dirac operator 
can reproduce various types of the Dirac operator index, 
including those with boundaries and the mod-two version.
Here we work on a two-dimensional square lattice
with size $L=33$ on which we assign $U(1)$ gauge link variables.
We present the results for the APS index on a two-dimensional disk $D^2$, 
mod-two APS index on $D^2$,  and the mod-two APS index on 
a torus from which a $D^2$ is hollowed out to form an $S^1$ boundary.
More comprehensive analysis was presented in \cite{Aoki:2025gca}.

We consider a circular domain-wall with radius $r_0=10$. 
In the region $\sqrt{x^2+y^2}<r_0$ we change the mass term $-sM$ 
by the one parameter $s\in [-1,1]$, while it is kept at $+M$ outside the domain-wall.
On the internal disk, we assign a $U(1)$ gauge potential given by
\begin{align}
 A_x(x,y)&=\left\{
\begin{array}{cc}
 -\frac{Q'y}{r_1^2}& (\sqrt{x^2+y^2}<r_1) \\
 -\frac{Q'y}{x^2+y^2} & (\sqrt{x^2+y^2}\ge r_1)
\end{array}
\right.,\;\;\;
 A_y(x,y)=\left\{
\begin{array}{cc}
 \frac{Q'x}{r_1^2}& (\sqrt{x^2+y^2}<r_1) \\
 \frac{Q'x}{x^2+y^2}& (\sqrt{x^2+y^2}\ge r_1)
\end{array}
\right.,
\end{align}
where $r_1=6$ is a fixed radius in which 
the constant curvature $2Q'/r_1^2$ is given
where $Q'$ equals to the total flux.
Then the link variables are set by
\begin{align}
U_x(x,y)&=\exp \left[i\int_{x}^{x+a} dx' A_x(x',y)\right],\;\;\;
U_y(x,y)=\exp \left[i\int_{y}^{y+a} dy' A_y(x,y')\right].
\end{align}

\begin{figure}[thbp]
\centering
  \includegraphics[width=0.7\columnwidth]{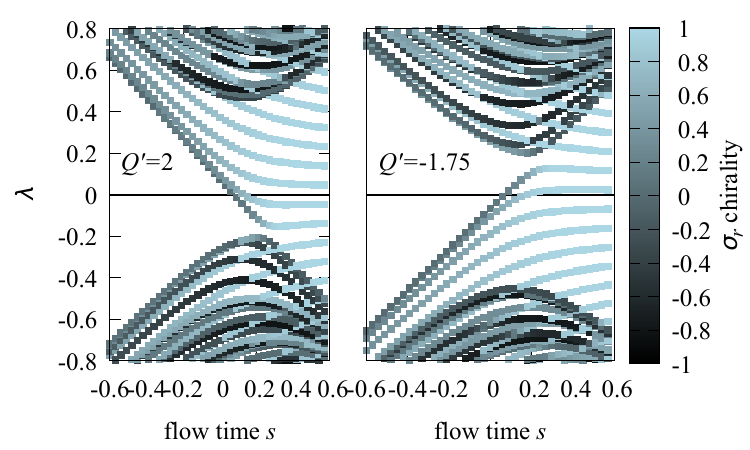}
  \caption{
    The eigenvalue spectrum the domain-wall Dirac operator
with a $U(1)$ flux $Q'=2$ (left panel) and $Q'=-1.75$ (right)
inside a circular domain-wall at $r=10$.
We assign $\epsilon=+1$ for $r<10$ (and $\epsilon=-1$ for $r\ge 10$).
In the both cases the APS index theorem on a disk is well reproduced.
  }
  \label{fig:APScircle}
  \end{figure}

In Fig.~\ref{fig:APScircle}, we plot the spectrum of the circular domain-wall Dirac operator
in the case with $Q'=2$ (left panel) and $Q'=-1.75$ (right).
The gradation denotes the expectation value of $\sigma_r$,
    which corresponds to the chirality operator of the edge-localized
    modes on the circle.
For $t>0$, the spectrum near zero becomes dense, which indicates
appearance of the one-dimensional massless modes on the edge.
From the dense mode spectrum at $t=1$, we can estimate
the values of the $\eta$ invariant of the boundary Dirac operator $iD^\mathrm{1D}$
whose spectrum is determined by the induced ``gravitational'' effect from the curved domain-wall
and Aharonov-Bohm effect of the $U(1)$ gauge field as shown in \cite{Aoki:2022aez}.
The estimated values are
$-\eta(iD^\mathrm{1D})/2=0$ for $Q'=2$, while 
 $-\eta(iD^\mathrm{1D})/2=-0.25$ for $Q'=-1.75$.
We observe that the both cases are consistent with the APS index theorem
\begin{equation}
 \mathrm{sf}\left[\gamma\left(D_W-\frac{s+1}{2} m \epsilon -\frac{s-1}{2}m\right)\right]= \underbrace{\frac{1}{2\pi}\int F}_{=Q'} -\frac{1}{2}\eta(iD^{1D}).
\end{equation}

Next we consider a free fermion case where the Dirac operator
\begin{equation}
 A_s = \sigma_1 \partial_x + \sigma_3 \partial_y + (i\sigma_2)\left[-\frac{s+1}{2} m \epsilon -\frac{s-1}{2}m\right], 
\end{equation}
is real. The corresponding lattice domain-wall Dirac operator is also real.
Here we assign the periodic boundary condition in both of the
    $x$ and $y$ directions.
For the domain-wall mass term, we try two cases.
One is the same as the previous case, giving 
$\epsilon=+1$ for $r<10$ (and $\epsilon=-1$ for $r\ge 10$),
with which the spectral flow should reproduce the mod-two index $=0$ on a disk.
The other is to give the opposite signs: 
$\epsilon=-1$ for $r<10$ (and $\epsilon=+1$ for $r\ge 10$),
which should lead to the mod-two index $=1$ on a torus $T^2$ with an $S^1$ boundary.

\begin{figure}[tbhp]
\includegraphics[width=0.7\columnwidth]{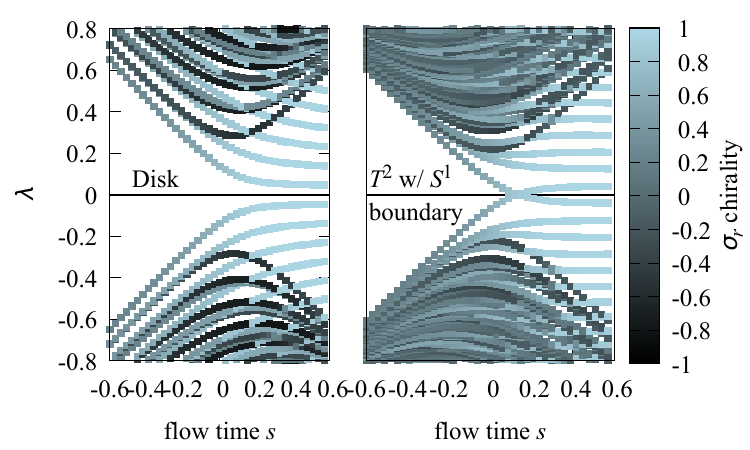}
\centering
  \caption{
    The eigenvalue spectrum of the free domain-wall Dirac operator with periodic boundary conditions.
    For the left panel, the domain-wall mass term is assigned 
    as $\epsilon=+1$ for $r<10$ (and $\epsilon=-1$ for $r\ge 10$)(Disk).
    For the right-panel, we set $\epsilon=-1$ for $r<10$ (and $\epsilon=+1$ for $r\ge 10$)
  ($T^2$ with $S^1$ boundary).
  }
  \label{fig:mod2APS}
\end{figure}

As presented in Fig.~\ref{fig:mod2APS}, 
the disk case (left panel) has no zero-crossings, while
in the $T^2$ with $S^1$ boundary case exhibits
one pair of zero-mode crossing.
Thus, the mod-two APS index is reproduced.

\section{Summary}

The massive Wilson Dirac operator can be identified as a mathematical object in $K$-theory,
giving an element in the group $K^1(I,\partial I)$ (or $KO^0(I,\partial I)$),  
and the associated spectral flow describes various index formulas. 

In our formulation, 
the chiral symmetry (GW relation) is not necessary at all, while
it perfectly agrees with the overlap Dirac index on periodic lattices.
Boundaries can be introduced by domain-walls,
which can be flat or curved.
For the latter, we can induce a gravitational background field.
Our formulation can be applied in arbitrary dimensions,
since the standard/mod-two versions are treated in a unified way.

The authors thank Yoshio Kikukawa and Yosuke Kubota
for helpful discussions.
This work was supported in part by JSPS KAKENHI Grant
(JP21K03222, JP21K03574, JP23K03387, JP23K22490, JP24K06719, JP25K07283, JP25K1738).
The work of SA was supported in part by RIKEN Special Postdoctoral Researchers Program.

\bibliographystyle{JHEP}
\bibliography{ref.bib}

@article{Aoki:2026mgx,
    author = "Aoki, Shoto and Fujita, Hajime and Fukaya, Hidenori and Furuta, Mikio and Matsuo, Shinichiroh and Onogi, Tetsuya and Yamaguchi, Satoshi",
    title = "{Capturing the Atiyah-Patodi-Singer index from the lattice}",
    eprint = "2602.12576",
    archivePrefix = "arXiv",
    primaryClass = "math.DG",
    reportNumber = "OU-HET-1300",
    month = "2",
    year = "2026"
}

@article{Aoki:2025gca,
    author = {Aoki, Shoto and Fukaya, Hidenori and Furuta, Mikio and Matsuo, Shinichiroh and Onogi, Tetsuya and Yamaguchi, Satoshi},
    title = {K-Theoretic Computation of the Atiyah(–Patodi)–Singer Index of Lattice Dirac Operators},
    journal = {Progress of Theoretical and Experimental Physics},
    volume = {2025},
    number = {6},
    pages = {063B09},
    year = {2025},
    month = {06},
    eprint = {2503.23921},
    issn = {2050-3911},
    doi = {10.1093/ptep/ptaf087},
}

@article{Aoki:2024sjc,
    author = "Aoki, Shoto and Fukaya, Hidenori and Furuta, Mikio and Matsuo, Shinichiroh and Onogi, Tetsuya and Yamaguchi, Satoshi",
    title = "{The index of lattice Dirac operators and $K$-theory}",
    eprint = "2407.17708",
    archivePrefix = "arXiv",
    primaryClass = "math.KT",
    reportNumber = "OH-HET-1236",
    month = "7",
    year = "2024"
}

@article{Fukaya:2021sea,
    author = "Fukaya, Hidenori",
    title = "{Understanding the index theorems with massive fermions}",
    eprint = "2109.11147",
    archivePrefix = "arXiv",
    primaryClass = "hep-th",
    reportNumber = "OU-HET-1106",
    doi = "10.1142/S0217751X21300155",
    journal = "Int. J. Mod. Phys. A",
    volume = "36",
    number = "26",
    pages = "2130015",
    year = "2021"
}

@article{Fukaya:2020tjk,
    author = "Fukaya, Hidenori and Furuta, Mikio and Matsuki, Yoshiyuki and Matsuo, Shinichiroh and Onogi, Tetsuya and Yamaguchi, Satoshi and Yamashita, Mayuko",
    title = "{Mod-two APS index and domain-wall fermion}",
    eprint = "2012.03543",
    archivePrefix = "arXiv",
    primaryClass = "hep-th",
    reportNumber = "OU-HET-1073",
    doi = "10.1007/s11005-022-01509-2",
    journal = "Lett. Math. Phys.",
    volume = "112",
    number = "2",
    pages = "16",
    year = "2022"
}

@article {MR4179728,
    AUTHOR = {Fukaya, Hidenori and Furuta, Mikio and Matsuo, Shinichiroh and
              Onogi, Tetsuya and Yamaguchi, Satoshi and Yamashita, Mayuko},
     TITLE = {The {A}tiyah-{P}atodi-{S}inger index and domain-wall fermion
              {D}irac operators},
   JOURNAL = {Comm. Math. Phys.},
  FJOURNAL = {Communications in Mathematical Physics},
    VOLUME = {380},
      YEAR = {2020},
    NUMBER = {3},
     PAGES = {1295--1311},
      ISSN = {0010-3616,1432-0916},
   MRCLASS = {58J20},
  MRNUMBER = {4179728},
MRREVIEWER = {Andres\ Larrain-Hubach},
       DOI = {10.1007/s00220-020-03806-0},
       URL = {https://doi.org/10.1007/s00220-020-03806-0},
}

@article{Fukaya:2017tsq,
    author = "Fukaya, Hidenori and Onogi, Tetsuya and Yamaguchi, Satoshi",
    title = "{Atiyah-Patodi-Singer index from the domain-wall fermion Dirac operator}",
    eprint = "1710.03379",
    archivePrefix = "arXiv",
    primaryClass = "hep-th",
    reportNumber = "OU-HET-946",
    doi = "10.1103/PhysRevD.96.125004",
    journal = "Phys. Rev. D",
    volume = "96",
    number = "12",
    pages = "125004",
    year = "2017"
}

@article {MR397797,
    AUTHOR = {Atiyah, M. F. and Patodi, V. K. and Singer, I. M.},
     TITLE = {Spectral asymmetry and {R}iemannian geometry. {I}},
   JOURNAL = {Math. Proc. Cambridge Philos. Soc.},
  FJOURNAL = {Mathematical Proceedings of the Cambridge Philosophical
              Society},
    VOLUME = {77},
      YEAR = {1975},
     PAGES = {43--69},
      ISSN = {0305-0041,1469-8064},
   MRCLASS = {58G10 (57D85 57E15)},
  MRNUMBER = {397797},
MRREVIEWER = {Kh.\ Knapp},
       DOI = {10.1017/S0305004100049410},
       URL = {https://doi.org/10.1017/S0305004100049410},
}

@article{Atiyah:1968mp,
    author = "Atiyah, M. F. and Singer, I. M.",
    title = "{The Index of elliptic operators. 1}",
    doi = "10.2307/1970715",
    journal = "Annals Math.",
    volume = "87",
    pages = "484--530",
    year = "1968"
}

@article{Neuberger:1997fp,
    author = "Neuberger, Herbert",
    title = "{Exactly massless quarks on the lattice}",
    eprint = "hep-lat/9707022",
    archivePrefix = "arXiv",
    reportNumber = "RU-97-63",
    doi = "10.1016/S0370-2693(97)01368-3",
    journal = "Phys. Lett. B",
    volume = "417",
    pages = "141--144",
    year = "1998"
}

@article{Hasenfratz:1998ri,
    author = "Hasenfratz, Peter and Laliena, Victor and Niedermayer, Ferenc",
    title = "{The Index theorem in QCD with a finite cutoff}",
    eprint = "hep-lat/9801021",
    archivePrefix = "arXiv",
    reportNumber = "BUTP-98-1",
    doi = "10.1016/S0370-2693(98)00315-3",
    journal = "Phys. Lett. B",
    volume = "427",
    pages = "125--131",
    year = "1998"
}

@article{Ginsparg:1981bj,
    author = "Ginsparg, Paul H. and Wilson, Kenneth G.",
    title = "{A Remnant of Chiral Symmetry on the Lattice}",
    reportNumber = "CLNS-81-520, HUTP-81-A060",
    doi = "10.1103/PhysRevD.25.2649",
    journal = "Phys. Rev. D",
    volume = "25",
    pages = "2649",
    year = "1982"
}

@article{Luscher:1998pqa,
    author = "Luscher, Martin",
    title = "{Exact chiral symmetry on the lattice and the Ginsparg-Wilson relation}",
    eprint = "hep-lat/9802011",
    archivePrefix = "arXiv",
    reportNumber = "DESY-98-014",
    doi = "10.1016/S0370-2693(98)00423-7",
    journal = "Phys. Lett. B",
    volume = "428",
    pages = "342--345",
    year = "1998"
}

@article{Itoh:1987iy,
    author = "Itoh, S. and Iwasaki, Y. and Yoshie, T.",
    title = "{The U(1) Problem and Topological Excitations on a Lattice}",
    reportNumber = "UTHEP-163",
    doi = "10.1103/PhysRevD.36.527",
    journal = "Phys. Rev. D",
    volume = "36",
    pages = "527",
    year = "1987"
}

@article{Adams:1998eg,
    author = "Adams, David H.",
    title = "{Axial anomaly and topological charge in lattice gauge theory with overlap Dirac operator}",
    eprint = "hep-lat/9812003",
    archivePrefix = "arXiv",
    doi = "10.1006/aphy.2001.6209",
    journal = "Annals Phys.",
    volume = "296",
    pages = "131--151",
    year = "2002"
}

@article{Clancy:2023ino,
    author = "Clancy, Michael and Kaplan, David B. and Singh, Hersh",
    title = "{Generalized Ginsparg-Wilson relations}",
    eprint = "2309.08542",
    archivePrefix = "arXiv",
    primaryClass = "hep-lat",
    reportNumber = "INT-PUB-23-024, IQuS@UW-21-064, FERMILAB-PUB-23-541-T",
    doi = "10.1103/PhysRevD.109.014502",
    journal = "Phys. Rev. D",
    volume = "109",
    number = "1",
    pages = "014502",
    year = "2024"
}

@article {MR4407739,
    AUTHOR = {Kubota, Yosuke},
     TITLE = {The index theorem of lattice {W}ilson-{D}irac operators via
              higher index theory},
   JOURNAL = {Ann. Henri Poincar\'e},
  FJOURNAL = {Annales Henri Poincar\'e. A Journal of Theoretical and
              Mathematical Physics},
    VOLUME = {23},
      YEAR = {2022},
    NUMBER = {4},
     PAGES = {1297--1319},
      ISSN = {1424-0637,1424-0661},
   MRCLASS = {19K56 (46L80 81T13)},
  MRNUMBER = {4407739},
MRREVIEWER = {Evgeniy\ V.\ Troitski\u i},
       DOI = {10.1007/s00023-022-01159-z},
       URL = {https://doi.org/10.1007/s00023-022-01159-z},
}

@article {MR4275791,
    AUTHOR = {Yamashita, Mayuko},
     TITLE = {A lattice version of the {A}tiyah-{S}inger index theorem},
   JOURNAL = {Comm. Math. Phys.},
  FJOURNAL = {Communications in Mathematical Physics},
    VOLUME = {385},
      YEAR = {2021},
    NUMBER = {1},
     PAGES = {495--520},
      ISSN = {0010-3616,1432-0916},
   MRCLASS = {58J20 (19K56)},
  MRNUMBER = {4275791},
MRREVIEWER = {Peter\ Haskell},
       DOI = {10.1007/s00220-021-04021-1},
       URL = {https://doi.org/10.1007/s00220-021-04021-1},
}

@article{Araki:2025xly,
    author = "Araki, Sho and Fukaya, Hidenori and Onogi, Tetsuya and Yamaguchi, Satoshi",
    title = "{The Arf-Brown-Kervaire invariant on a lattice}",
    eprint = "2512.11424",
    archivePrefix = "arXiv",
    primaryClass = "hep-lat",
    reportNumber = "OU-HET-1294",
    month = "12",
    year = "2025"
}

@unpublished{Araki,
    author = "Araki, Sho and Fukaya, Hidenori and Onogi, Tetsuya and Yamaguchi, Satoshi",
    eprint = " ",
    note = "in preparation",
    year = "in preparation", 
}

@article{Nguyen:2024wck,
    author = "Nguyen, Mendel and Singh, Hersh",
    title = "{Chiral symmetry and Atiyah-Patodi-Singer index theorem for staggered fermions}",
    eprint = "2405.11348",
    archivePrefix = "arXiv",
    primaryClass = "hep-lat",
    reportNumber = "FERMILAB-PUB-24-0262-T",
    month = "5",
    year = "2024"
}

@article{Carey2016SpectralFF,
  title={Spectral flow for skew-adjoint Fredholm operators},
  author={Alan Carey and John Phillips and Hermann Schulz-Baldes},
  journal={Journal of Spectral Theory},
  year={2016},
  url={https://api.semanticscholar.org/CorpusID:119154998}
}

@article{Kaplan:1992bt,
    author = "Kaplan, David B.",
    title = "{A Method for simulating chiral fermions on the lattice}",
    eprint = "hep-lat/9206013",
    archivePrefix = "arXiv",
    reportNumber = "UCSD-PTH-92-16",
    doi = "10.1016/0370-2693(92)91112-M",
    journal = "Phys. Lett. B",
    volume = "288",
    pages = "342--347",
    year = "1992"
}

@article{Shamir:1993zy,
    author = "Shamir, Yigal",
    title = "{Chiral fermions from lattice boundaries}",
    eprint = "hep-lat/9303005",
    archivePrefix = "arXiv",
    reportNumber = "WIS-93-20-PH",
    doi = "10.1016/0550-3213(93)90162-I",
    journal = "Nucl. Phys. B",
    volume = "406",
    pages = "90--106",
    year = "1993"
}

@article{Fukaya:2019myi,
    author = "Fukaya, Hidenori and Kawai, Naoki and Matsuki, Yoshiyuki and Mori, Makito and Nakayama, Katsumasa and Onogi, Tetsuya and Yamaguchi, Satoshi",
    title = "{The Atiyah{\textendash}Patodi{\textendash}Singer index on a lattice}",
    eprint = "1910.09675",
    archivePrefix = "arXiv",
    primaryClass = "hep-lat",
    reportNumber = "OU-HET-1027",
    doi = "10.1093/ptep/ptaa031",
    journal = "PTEP",
    volume = "2020",
    number = "4",
    pages = "043B04",
    year = "2020"
}

@article{Aoki:2022aez,
    author = "Aoki, Shoto and Fukaya, Hidenori",
    title = "{Curved domain-wall fermion and its anomaly inflow}",
    eprint = "2212.11583",
    archivePrefix = "arXiv",
    primaryClass = "hep-lat",
    reportNumber = "OU-HET-1164",
    doi = "10.1093/ptep/ptad023",
    journal = "PTEP",
    volume = "2023",
    number = "3",
    pages = "033B05",
    year = "2023"
}

@article{Luscher:2006df,
    author = "Luscher, Martin",
    title = "{The Schrodinger functional in lattice QCD with exact chiral symmetry}",
    eprint = "hep-lat/0603029",
    archivePrefix = "arXiv",
    reportNumber = "CERN-PH-TH-2006-052",
    doi = "10.1088/1126-6708/2006/05/042",
    journal = "JHEP",
    volume = "05",
    pages = "042",
    year = "2006"
}
\end{document}